\documentclass[%
 reprint,
 amsmath,amssymb,
 aps,
floatfix,
]{revtex4-1}

\usepackage{graphicx}% Include figure files
\usepackage{dcolumn}% Align table columns on decimal point
\usepackage{bm}% bold math
\begin{document}

%\preprint{APS/123-QED}

\title{Robust entropy requires strong and balanced excitatory and inhibitory synapses}

\author{Vidit Agrawal\textsuperscript{1}}
\thanks{These authors contributed equally to this work.}
\author{Andrew B. Cowley\textsuperscript{2}}
\thanks{These authors contributed equally to this work.}
\author{Qusay Alfaori$^1$}%
\author{Juan G. Restrepo$^2$}
\author{Daniel B. Larremore$^{3,4}$}
\author{Woodrow L. Shew$^1$}%
\email{shew@uark.edu}

\affiliation{%
$^1$Department of Physics, University of Arkansas, Fayetteville, Arkansas, USA\\
$^2$Department of Applied Mathematics, University of Colorado, Boulder, Colorado, USA\\
$^3$Department of Computer Science, University of Colorado, Boulder, Colorado, USA \\
$^4$BioFrontiers Institute, University of Colorado, Boulder, Colorado, USA
}



\begin{abstract}
It is widely appreciated that well-balanced excitation and inhibition are necessary for proper function in neural networks. However, in principle, such balance could be achieved by many possible configurations of excitatory and inhibitory strengths, and relative numbers of excitatory and inhibitory neurons. For instance, a given level of excitation could be balanced by either numerous inhibitory neurons with weak synapses, or few inhibitory neurons with strong synapses. Among the continuum of different but balanced configurations, why should any particular configuration be favored? Here we address this question in the context of the entropy of network dynamics by studying an analytically tractable network of binary neurons.  We find that entropy is highest at the boundary between excitation-dominant and inhibition-dominant regimes.  Entropy also varies along this boundary with a trade-off between high and robust entropy: weak synapse strengths yield high network entropy which is fragile to parameter variations, while strong synapse strengths yield a lower, but more robust, network entropy. 
In the case where inhibitory and excitatory synapses are constrained to have similar strength, we find that a small, but non-zero fraction of inhibitory neurons, like that seen in mammalian cortex, results in robust and relatively high entropy. 
%\begin{description}

%\end{description}
\end{abstract}

\pacs{Valid PACS appear here}% PACS, the Physics and Astronomy
\maketitle

%\linenumbers

\section*{Introduction}

The network of neurons in cerebral cortex displays rich and complex dynamics even when not engaged by any particular sensory or motor interaction with the external world \cite{fox2007, arieli1996}.  From one point of view, such ongoing internal dynamics are thought to mediate memory consolidation and other internal cognitive processes \cite{Luczak2009,Han2008,berkes2011, romano2015, miller2014}.  On the other hand, ongoing fluctuations in cortical network dynamics have often been considered a nuisance, imposing noisy fluctuations in neural response to sensory input \cite{ecker2014, lee1998, averbeck2006}.  In both of these contexts, it is important to understand the mechanisms which govern the fluctuations of ongoing cortical network dynamics.  Here we investigate Shannon entropy of macroscopic network dynamics. In the context of internal cognitive processes, high entropy might be beneficial, corresponding to a larger repertoire of internal states to mediate internal information transfer \cite{fagerholm2016}.  When considered as noise, high entropy can be a hindrance to effective sensory coding.  Indeed, in principle, encoding of sensory input would be most reliable if the cortex was totally silent (low entropy) until the stimulus excited it.  However, real cortex does not operate this way; it has many jobs to do beyond encoding sensory input and is never silent.  Previous studies have shown that ongoing cortical dynamics with high entropy occurs together with high mutual information between stimulus and response \cite{shew2011, fagerholm2016}, suggesting that a large repertoire of ongoing dynamical states may be necessary for a large repertoire of stimulus-evoked states \cite{Luczak2009, berkes2011}.

A crucial factor for determining the entropy of network dynamics in the cortex is the competition between two types of neurons: excitatory (E) and inhibitory (I).  The importance of balanced excitation and inhibition is most apparent in previous experiments that directly manipulated the E/I balance pharmacologically.  These studies have shown that ongoing network dynamics can vary dramatically when GABA synapses are either enhanced or suppressed \cite{mao2001,shew2011,fagerholm2016, gautam}. Enhanced inhibition (GABA agonists) often results in a dynamical regime characterized by low firing rates and weak population-level correlations, while decreased inhibition (GABA antagonists) tends to result in a regime with higher firing rates and strong correlations. Two studies in particular have shown that entropy can be increased by tuning the E/I balance to the tipping point between these two distinct dynamical regimes \cite{shew2011,fagerholm2016}. However, more systematic understanding of how E/I balance impacts entropy is difficult to obtain experimentally because pharmacological manipulations are rather difficult to precisely control. Moreover, with a few interesting exceptions \cite{chen2010,hunt2013}, experiments do not vary the numbers of excitatory or inhibitory neurons. Computational models offer an alternative approach in which the number of excitatory and inhibitory neurons, as well as strength of excitatory and inhibitory synapses, can easily be controlled.  A few previous computational studies have addressed similar topics, but typically have neglected inhibition \cite{shew2011, ferraz} or have not considered the effects of changing the E/I ratio \cite{scarpetta2013,zhou}. 
Thus, theoretical and experimental understanding of the relationship between the entropy of ongoing dynamics and the balance of excitation and inhibition---mediated by both relative strengths of excitatory and inhibitory synapses and relative numbers of excitatory and inhibitory cells---remains unresolved.

Here we attempt to improve the theoretical understanding of entropy of ongoing dynamics by studying a network model of binary neurons in detail.  We consider how network entropy depends on the fraction of inhibitory neurons $\alpha$ and the strengths of E and I interactions, $W_E$ and $W_I$.  We find maximal entropy near the tipping point between the low and high firing rate dynamical regimes, as seen in experiments.  For a given choice of $W_E$ and $W_I$, we find that the tipping point can be achieved by adjusting the value of $\alpha$; this raises the question of why any particular configuration of parameters should be favored over another.
We find that there is a trade-off between high and robust network entropy: networks with weak synapses can achieve a high entropy when excitation and inhibition are balanced, but the entropy degrades significantly upon small deviations from the balanced state. On the other hand, networks with stronger synapses have a lower optimal entropy, but they are more robust to parameter changes. We also find that if E and I synaptic strengths are proportional to each other, as found in many experiments \cite{deneve, wehr, haider}, then robust, high entropy requires a small  fraction of I neurons ($\alpha$ of order 0.1). In mammalian cortex, $\alpha$ has been found to be near 0.2 with remarkable consistency over the lifetime of an organism \cite{sahara} and over different regions of cortex \cite{hendry, meinecke}. Our results suggest that mammalian cortex strikes a compromise with intermediate, but robust entropy.  

In what follows, we introduce and analyze the binary neuron model which both predicts and provides insight into the results of model numerical simulations.

\section*{Model and theory}\label{binarymodel}

\subsection{Binary neuron model}\label{binary}
We explore the effects of excitation and inhibition balance on entropy using a simple, analytically tractable model.  The model, studied previously in Ref.~\cite{larremore2014}, consists of a network of $N$ stochastic binary neurons, indexed $i = 1,2,\dots, N$. The state of neuron $i$ at time $t$ is denoted by $x_i^t$, which can take the values $x_i^t = 0$ if the neuron is resting and $x_i^t = 1$ if the neuron is spiking. Time is assumed to evolve in discrete steps $t = 0,1,2,\dots$. 
The evolution of each neuron's state is stochastic and depends on the states of other neurons at the previous time step. It is given by
\begin{align}\label{xs}
  x_i^{t+1} = \left\{ \begin{array}{ll }
       1 & \text{with probability } \eta + (1-\eta)\sigma\left(\sum_{j=1}^N \epsilon_j w_{ij} x_j^t \right),\\
       0 & \text{otherwise,} 
  \end{array}\right.
\end{align}
 where $\epsilon_j = 1$ ($\epsilon_j = -1$) if neuron $j$ is excitatory (inhibotory), $w_{ij} > 0$ is the strength of the synapse  from neuron $j$ to neuron $i$ (which is taken to be zero if neuron $j$ does not connect to neuron $i$), and $\sigma(x) = \min(1, \max(0,x))$ is a transfer function that converts the input to neuron $i$ into a probability. The constant $\eta = 1/(100 N)$ represents independent spontaneous activation due to external sources, resulting in one spike per 100 time steps among all neurons, on average.  We consider Erd\H{o}s-R\'enyi networks where a directed link is made independently from neuron $j$ to neuron $i$ with probability $k/(N-1)$ for all $i \neq j$. The parameter $k$ is the expected number of outgoing connections from a given neuron. To control the relative number of excitatory and inhibitory neurons, we assign each neuron to be inhibitory with probability $\alpha$ and excitatory otherwise. Finally, we assume for simplicity that $w_{ij} = w_E$ for excitatory synapses (i.e., if $\epsilon_j = 1$) and  $w_{ij} = w_I$ for inhibitory synapses (i.e., if $\epsilon_j = -1$), and define the effective excitatory weight as $W_E = k w_E$ and the effective inhibitory weight as $W_I$ = $k w_I$.

The model is characterized by the parameters $N$, $k$, $W_E$, $W_I$, and $\alpha$. For definiteness, in the rest of the paper we will consider, unless otherwise indicated, only the parameters $N = 10000$ and $k = 100$, and study the macroscopic dynamics of the model as a function of $(W_E, W_I, \alpha)$. As a measure of collective network dynamics we study the fraction of spiking neurons, or {\it network activity}, given by
\begin{align}
    S^t = \frac{1}{N} \sum_{i=1}^N x^t_i.
    \label{eq:activity}
\end{align}
\begin{figure}[t]
\includegraphics{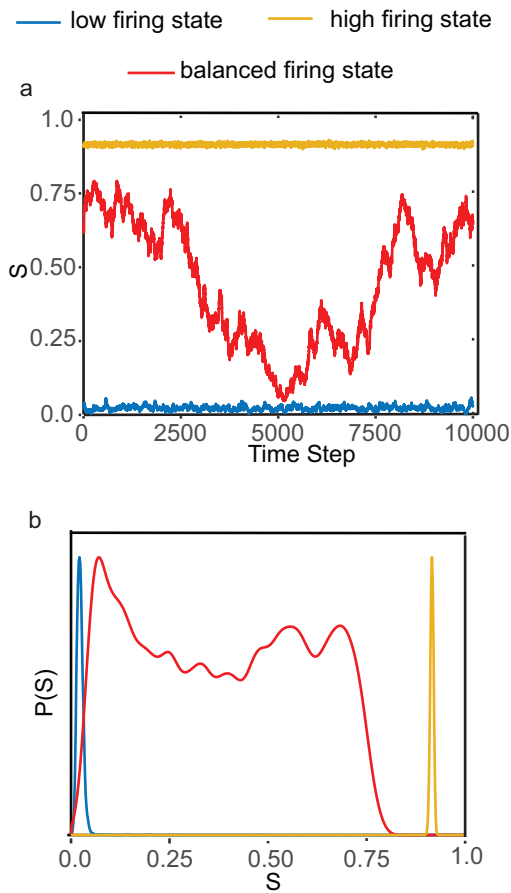}
\renewcommand{\figurename}{Fig}
\caption{{\bf Network activity and dynamics of binary model.}
Time series of network activity (a) show diverse fluctuations when excitation and inhibition are balanced ($\lambda=1$).  Similarly, probability distributions (b) of network activity are broadest when $\lambda=1$.  All probability distributions have been normalized by their peak probability to facilitate comparison of their shapes. Dynamical parameters: $\alpha$ = 0.11 (Blue), $0.1$ (Red), $0.09$ (Yellow); $W_E$=$W_I$=$1.25$.}
\label{fig1}
\end{figure}

In Ref.~\cite{larremore2014} it was found that the collective dynamics of the network is determined by the largest eigenvalue $\lambda$ of the connection strength matrix $A$ with entries $\{\epsilon_j w_{ij}\}_{i,j = 1}^N$. Network activity saturates at a high value for $\lambda > 1$ and dies out or reaches a steady low value for $\lambda < 1$. At the tipping point between these two regimes, defined by $\lambda = 1$, excitation and inhibition are balanced such that network activity is characterized by large fluctuations that are effectively ceaseless (their lifetime scales exponentially with $N$) \cite{larremore2014}. Figure~\ref{fig1}a shows an example of the time series of network activity for these three regimes.  For the Erd\H{o}s-R\'enyi networks considered here, $\lambda$ can be approximated by the expected row sum of $A$, 
\begin{align}\label{estimate}
\lambda \approx k w_E (1 - \alpha) - k w_I \alpha = W_E (1-\alpha) -W_I \alpha.
\end{align}
With this approximation, then, the parameters that give $\lambda = 1$ form a 2-dimensional surface in the $(W_E,W_I,\alpha)$ parameter space.

\subsection{Entropy}

We consider the Shannon entropy of the time-series of network activity, which quantifies the size of the repertoire of accessible macroscopic network states. The network activity is discrete (i.e., $0,1/N,2/N\dots,1)$. For a given set of network parameters, $(W_E,W_I,\alpha)$ 
we consider the steady-state probability distribution of network activity $P(S)$ and the associated entropy,  
\begin{align}\label{entropy}
H = -\sum_S P(S) \log_2(P(S)),
\end{align}
where the sum runs over the allowed values $S = 0, 1/N,2/N,\dots, 1$. In practice, we estimate $P(S)$ numerically from a time series of $S^t$ obtained from model simulations (Fig.~\ref{fig1}b) or from our semi-analytical theory, presented below, that treats the evolution of $S^t$ as a biased random walk.

\subsection{Simulation-free theory}

 Here we present a semi-analytical approach to compute the entropy for a given set of parameter values in the binary model. While numerical simulation of Eqs.~(\ref{xs}) for different values of $(W_E, W_I, \alpha)$ allows us to search for regimes yielding high and robust entropy, the complementary semi-analytical approach presented below is useful because it does not suffer from fluctuations caused by specific network realizations and, especially as its computational complexity does not depend on $N$, it can be faster than direct simulation of Eqs.~(\ref{xs}). Most importantly, our semi-analytical approach provides insights into our main findings.

The main idea of our approach is to treat the evolution of the macroscopic variable $S^t$ as a biased random walk. Although in principle the dynamics of the system depends on the microscopic states $\{x_n\}_{n=1}^N$, for large homogeneous networks one can describe the evolution of the system in terms of the macroscopic variable $S^t$.  
To analyze this random variable, one should determine if at any given time it is expected to decrease or increase.
This information is encapsulated in the {\it branching function} introduced in Ref~\cite{larremore2014} as the ratio $\Lambda(S) = E[S^{t+1}| S^t = S]/S$, where the expected value is taken over realizations of the stochastic dynamics and microscopic configurations with activity $S$.
In our case, the branching function can be approximated by \cite{larremore2014}, 
\begin{align}\label{Lambdaeq}
	\Lambda(S) = \frac{1}{S}E_P[\sigma( w_E n_E - w_I n_I)]\ ,
\end{align}
where the random variables $n_E$ and $n_I$ represent the number of active E and I inputs to a single neuron, respectively. Because we consider random networks, $n_E$ and $n_I$ are given by Poisson random variables with means $k S (1-\alpha)$  and $k S \alpha$, respectively.  The expected value $E_P[\cdot]$ is an expected value over the random variables $n_E$ and $n_I$. By assuming that the statistics of the macroscopic dynamics depend only on $S$, one can then write a random walk model for $S^t$ as
\begin{align}
	S^{t+1} = S^t\Lambda(S^t) + r(S^t)\ ,
\end{align}
where $r$ represents statistical noise which, by the definition of $\Lambda$, has mean zero. To obtain a tractable model we assume that $r(S^t)$ is normally distributed and has variance $V(S^t) = S^t(1-S^t)/N$, as estimated in Ref.~\cite{larremore2014}. This approximation is what one would obtain if each of the $N$ neurons is independently assumed to be active with probability $S$ and inactive with probability $1-S$. In this approximation, the probability that the system makes a transition from a state with activity $S'$ to a state with activity $S$ is given by
\begin{align}
	T(S|S') = \frac{1}{\sqrt{2 \pi V(S')}} \exp\left(-\frac{[S'\Lambda(S') - S]^2}{2 V(S')}\right)\ .
\end{align}
The distribution $P^t(S)$ of $S$ at time $t$ evolves following the master equation
\begin{align}\label{mapint}
	P^{t+1}(S) = \int_0^1  T(S|S') P^t(S') d S'\ ,
\end{align}
and as $t \to \infty$ it converges to a steady-state, which may be calculated numerically as the Perron-Frobenius eigenvector (with eigenvalue 1) of the linear operator 
\begin{align}\label{integral}
	\mathcal{L}\{P\}(S) = \int_0^1 T(S|S') P(S')  d S'\ .
\end{align}
The eigenvector can be calculated numerically by discretization of the integral in Eq.~(\ref{integral}) or as the limit of repeated iterations of Eq.~(\ref{mapint}). The entropy is then calculated directly from Eq. (\ref{entropy}).

\section*{Results}

Our primary goal is to determine how the entropy of a network varies with the relative numbers of E and I neurons and the relative strength of E and I synapses.   We first describe our results from numerical simulations of the binary model and then describe results from the theory.

First, we show in Fig.~1 that the system network activity visits the widest variety of states when excitation and inhibition are balanced at the tipping point between high and low firing rate regimes.  This is visible in time series
(Fig.~1a) as well as empirical distributions $P(S)$ of network activity (based on $10^4$ time steps of simulation).  Correspondingly, entropy $H$ is greatest along the boundary between low and high firing regimes (Fig.~2).  In the three-dimensional $(W_E,W_I, \alpha)$ parameter space this boundary forms a curved surface, which we henceforth refer to as the {\it maximum entropy surface}.  

As discussed in Sec.~\ref{binary}, we expect that the transition from the low to the high firing regimes occurs at the {\it critical surface} of parameters where $\lambda = 1$. While we find this is usually an excellent approximation to our numerical results, the maximum entropy and critical surfaces differ slightly for high values of $\alpha$, and therefore we will only use the critical surface as a qualitative guide to the location of the maximum entropy surface.   

To numerically identify the maximum entropy surface, for each fixed value of $(W_E, W_I)$ we compute entropy across a wide range of values of $\alpha$, finding the value $\alpha^*$ that maximizes $H(W_E,W_I,\alpha)$. In Fig.~\ref{fig3}a we show $\alpha^*$ as a function of {$W_E$ and $W_I$}. As one might expect, higher values of $W_E$ require a larger number of I neurons (higher $\alpha^*$) in order to maintain a balanced network, and vice versa. This agrees qualitatively with the estimate using the critical surface,  $\alpha^* \approx (W_E - 1)/(W_E + W_I)$ obtained from \eqref{estimate} with $\lambda = 1$.

\begin{figure*}[t]
\includegraphics{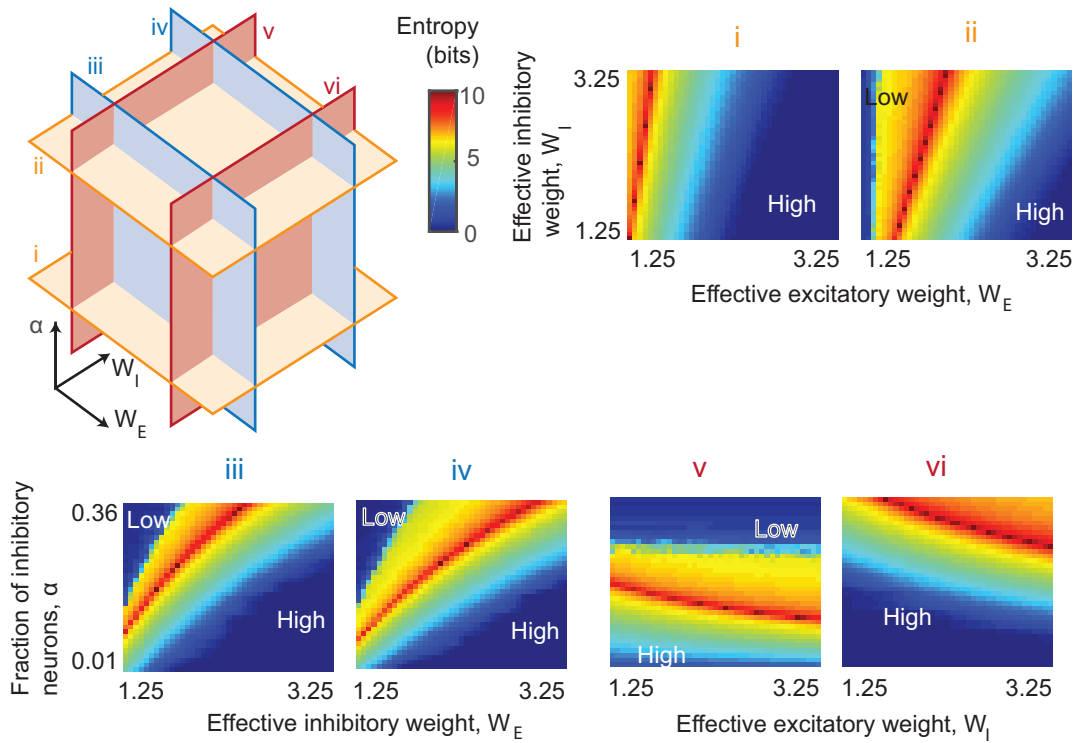}
\renewcommand{\figurename}{Fig}
\caption{{\bf  High entropy at boundary between high and low firing regimes.}
  Each panel shows how entropy (color) varies across a two-dimensional section of the three-dimensional $W_E$-$W_I$-$\alpha$ parameter space. Relative orientation of the six different sections are illustrated and labeled (i-iv) in the cartoon (left). For i and ii, $\alpha$ is fixed at $0.1$ and $0.2$. For iii and iv $W_I$ is fixed at $1.5$ and $2.5$. For v and vi $W_E$ is fixed at $1.5$ and $2.5$. A curved critical surface in $W_E$-$W_I$-$\alpha$ space separates the high firing regime (H) from a low firing regime (L). Entropy is high along this regime boundary.  Note that as I or E synapse strength increases the width of the peak in entropy also increases, indicating increased robustness (decreased fragility).
% Range for dynamical parameters $\alpha$: $0.01-0.36$, $W_E$: $1.25-3.25$ and $W_I$: $1.25-3.25$
}
\label{fig2}
\end{figure*}

Having identified the parameters that characterize the  maximum entropy surface, we next ask two questions.  First, where on the surface is entropy highest?  Second, where on the surface is entropy most robust?  We consider the entropy to be robust if it does not drop dramatically when we make a small perturbation in $W_E$, $W_I$, and $\alpha$ away from the peak entropy surface.  This approach is similar to other ways to quantify sensitivity to model parameters, such as Fisher information \cite{Lehmann1998}.  To quantify how much the entropy decreases if parameters are perturbed away from the maximum entropy surface, we define {\it fragility} $F(W_E,W_I)$ as follows.  For a given pair of $(W_E,W_I)$ values, we first calculate the entropy at the corresponding point on the maximum entropy surface, $H^* = H(W_E,W_I,\alpha^*)$. Then, we calculate the entropy at two points at a small distance $\delta$ above and below the surface, $H_{\text{up}}=H(W_E+\Delta W_E,W_I+\Delta W_I,\alpha+\Delta\alpha)$ and $H_{\text{down}}=H(W_E-\Delta W_E,W_I-\Delta W_I,\alpha-\Delta\alpha)$.  The perturbations $\pm(\Delta W_E,\Delta W_I,\Delta\alpha)$ are defined to be normal to the maximum entropy surface, which will give the largest drop in entropy for a given perturbation size.  The size of the perturbation was chosen to be small (Euclidean norm $\delta = 0.01$, about 1\% variation in parameters) because naturally occurring changes in E, I, and $\alpha$ are not likely to be large.  Finally, we define {\it fragility} $F(W_E,W_I)$ as the mean of the entropy difference;
\begin{align}
F(W_E,W_I) = \frac{(H^*-H_{\text{up}})+(H^*-H_{\text{down}})}{2} .
\end{align}

\begin{figure*}[t]
\includegraphics{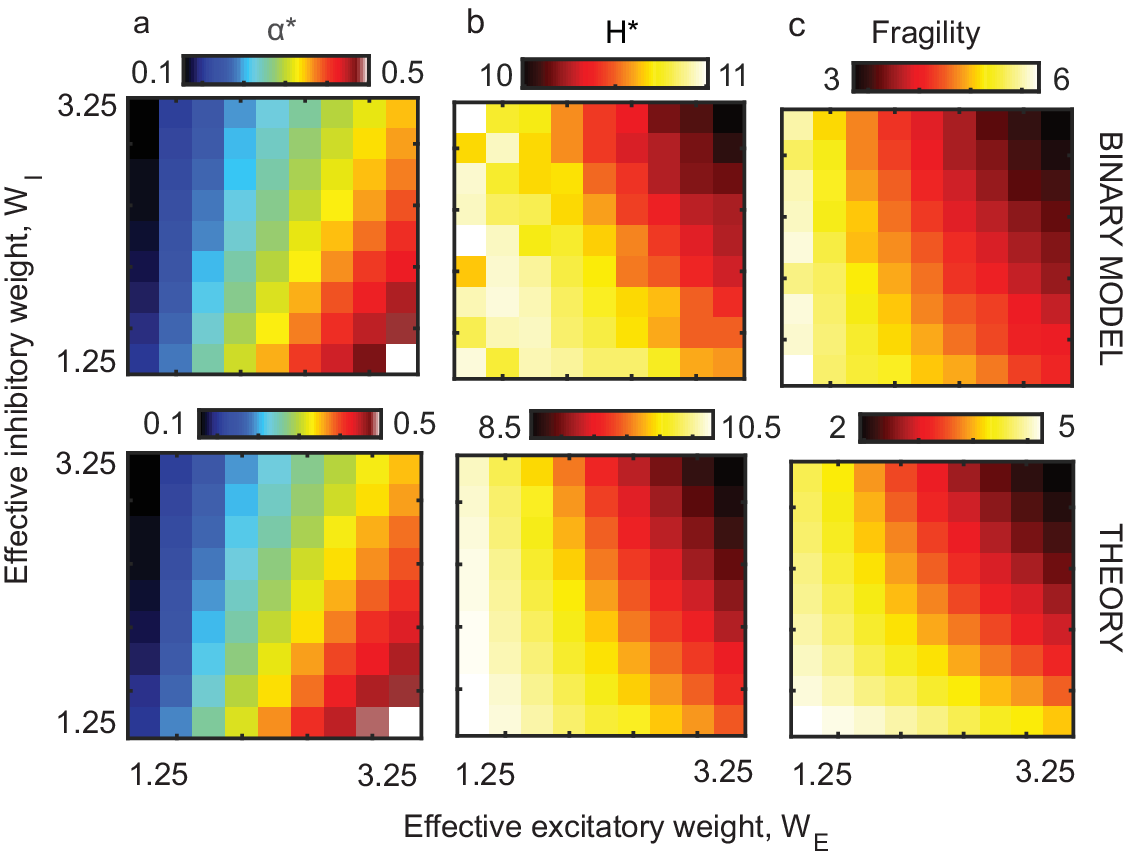}
\renewcommand{\figurename}{Fig}
\caption{{\bf Trade off between high entropy and robust entropy.}
a) For each combination of $W_E$ and $W_I$ effective synaptic weights, we identify the critical fraction of inhibitory neurons ($\alpha$*) with the highest entropy.  
b) Comparing all critical entropy $H^*$ across the entire critical surface, entropy was highest for low $W_E$ and $W_I$. c) Highest fragility was also found for low $W_E$ and $W_I$.}
\label{fig3}
\end{figure*}

Our main results are in Figs.~\ref{fig3}b and \ref{fig3}c. Figure~\ref{fig3}b shows the entropy $H^*$ on the maximum entropy surface as a function of the effective E and I weights $W_E$ and $W_I$. Networks with weak effective synapse strengths (low values of $W_E$ and $W_I$) can achieve a higher entropy $H^*$ than networks with strong effective synapse strengths. However, as shown in Fig.~\ref{fig3}c, high entropy comes at the cost of high fragility: networks with weak effective synapse strengths have the highest fragility, while networks with strong effective synapse strengths are the most robust. We note that while the variation in entropy $H^*$ is relatively moderate across the range studied (approximately $10\%$), the fragility ranges from $3$ to $6$, indicating that our $1\%$ perturbation of parameters results in a dramatic drop in entropy of approximately $30\%$ to $60\%$. One could argue that what matters are the final values of entropy after perturbation (i.e., $H_{up}$ and $H_{down}$) rather than how much entropy drops due to perturbation (i.e., $F$).  From this perspective, strong synapses are also better; $H_{up}$ and $H_{down}$ are lower for weak synapses than for strong synapses.  This can be seen by subtracting Fig.~\ref{fig3}c from \ref{fig3}b.  We conclude that there is a trade-off between high and robust entropy, with stronger effective synapse strengths promoting lower but more robust entropy, and weaker effective synapse strengths promoting a high but fragile entropy. 

Finally, we  address the role of the fraction  $\alpha$ of I neurons in promoting entropy robustness. We note that if the choices of E and I synapse strengths are constrained to be proportional to each other, as experiments suggest \cite{deneve, wehr, haider}, then $W=W_E =b W_I$ and the estimate $\alpha^* \approx (W_E - 1)/(W_E + W_I)$ becomes $\alpha^* = (1+1/b)^{-1} (1 - 1/W)$.  Thus, $\alpha^*$ is a monotonically increasing function of synapse strength. Therefore, for such constrained networks, entropy and fragility decreases with the fraction of I neurons $\alpha$. Thus, a small non-zero $\alpha$, similar to mammalian cortex, is needed to obtain high and robust entropy.

\begin{figure*}[t]
\includegraphics{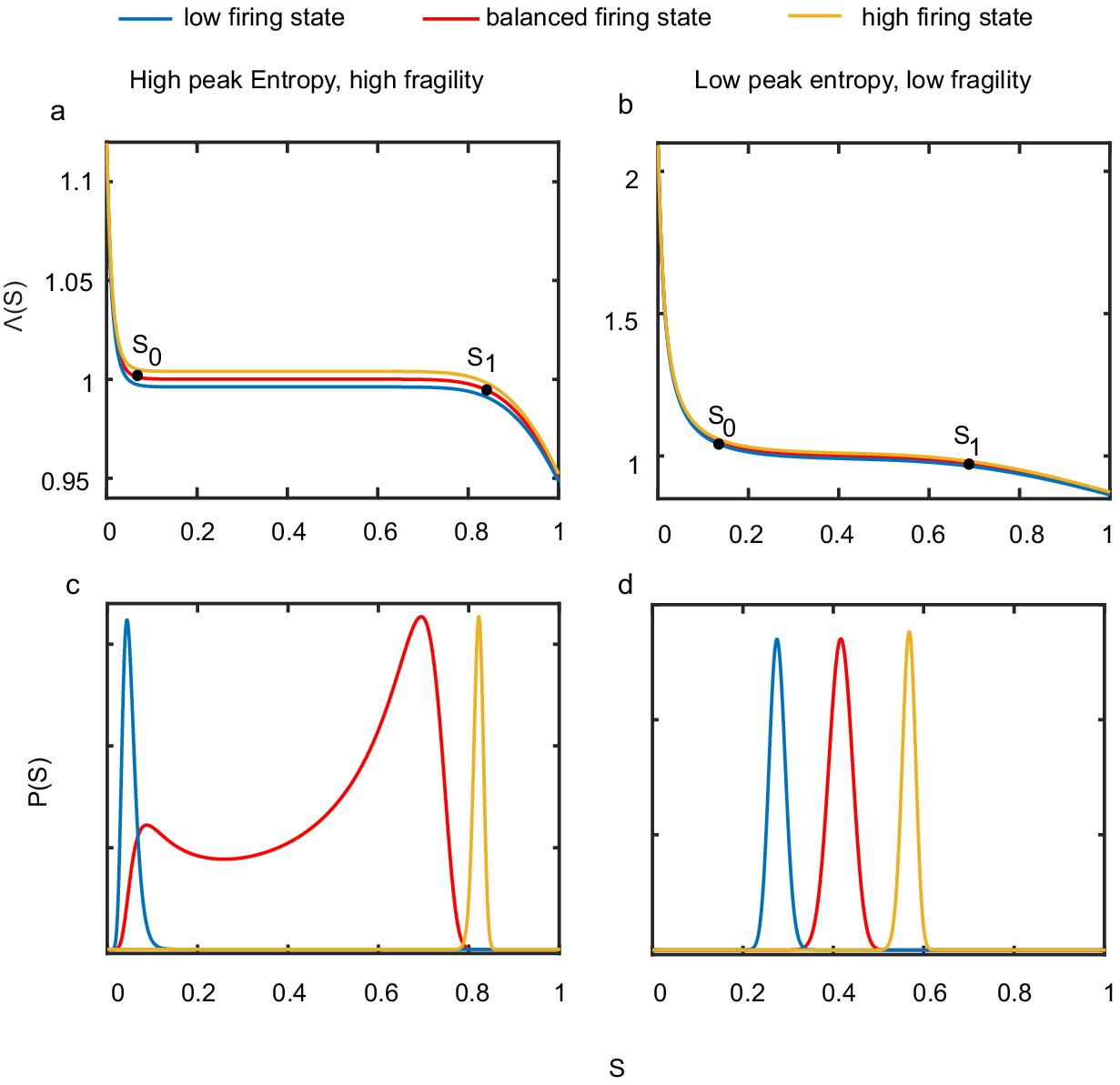}
\renewcommand{\figurename}{Fig}
\caption{{\bf Interpretation of results based on Branching function formalism.}
Branching functions $\Lambda(S)$, for a) low effective excitatory and inhibitory weight ($W_E = W_I= 1.25$) with $S_0 \sim 0.015$ and $S_1 \sim 0.883$, and b) high effective excitatory and inhibitory weight ($W_E = W_I= 3.25$) with $S_0 \sim 0.105$ and $S_1 \sim 0.724$. The probability distributions c) low effective weights and d) high effective weights. All probability distributions have been normalized by their peak probability to facilitate comparison of their shapes.}
\label{fig4}
\end{figure*}
In the following, we present an interpretation of our results based on the branching function formalism presented above and studied in previous work \cite{larremore2014}. If one treats the time series of network activity $S^t$ as a random walk, its bias, or expected velocity, is given by $S^t (\Lambda(S^t)-1)$. Therefore, when $\Lambda(S^t) > 1$ ($\Lambda(S^t) < 1$), $S^t$ tends to increase (decrease). Since $\Lambda(0) \geq 1$ and $\Lambda(1) \leq 1$ \cite{larremore2014}, the long-time distribution of $S^t$ will be concentrated around the region where $\Lambda(S^t) \approx 1$. The wider this region is, the wider the distribution of $S$ will be, and the larger its associated entropy. To understand how the size of this region depends on the weights $W_E$ and $W_I$, we note that, at the tipping point between low and high firing regimes, the branching function deviates from $1$ in an interval $[0,S_0)$ on which it is appreciably larger than $1$, and in an interval $(S_1, 1]$ on which it is less than $1$. The branching function deviates from $1$ in these intervals because the distribution of the random variable $w_E n_E - w_I n_I$ in Eq.~(\ref{Lambdaeq}) extends below $0$ or above $1$ when $S$ is too close to $0$ or $1$, respectively. In these cases, the nonlinearity in the transfer function $\sigma$ causes the expected value in Eq.~(\ref{Lambdaeq}) to be different from $1$. The larger the values of $w_E$ and $w_I$, the wider the distribution of $w_E n_E - w_I n_I$, and therefore the larger these intervals are. More precisely, we can estimate the scaling of $S_0$ and $S_1$ as follows. The variance of the variable $w_E n_E - w_I n_I$ is $V(S) =w_E^2(1-\alpha) k S + w_I^2 \alpha k S$. Estimating $S_0$ and $S_1$ as the values where $S_0^2 \sim V(S_0)$ and $(1-S_1)^2 \sim V(S_1)$ we obtain $S_0 \sim w_E^2(1-\alpha) k + w_I^2 \alpha k$ and $S_1 \sim 1-\sqrt{\left(\frac{1}{2}
   S_0+1\right)^2-1} + \frac{1}{2} S_0$. Using the approximation that in the balanced state $\alpha = (k w_E -1)/(k w_E + k w_I)$, this gives closed expressions for the estimates of $S_0$ and $S_1$ as a function of $w_E$ and $w_I$. For low values of $w_E$ and $w_I$, $S_0 \ll 1$ and $1-S_1 \ll 1$, and therefore the branching function will be close to $1$ over a large region in $[0,1]$. This is illustrated in the left panel of Fig.~\ref{fig4}, which shows the branching function $\Lambda(S)$ and associated probability distribution $P(S)$  for the balanced state (red lines), high-firing (yellow lines) and low-firing (blue lines) cases. While the wide region over which the branching function is approximately one results in a relatively large entropy, a perturbation away from the balanced state displaces the branching function so that it is below or above $1$ over this large region, and is close to $1$ over a much smaller region. Thus, the entropy decreases significantly. On the other hand, if the weights $w_E$ and $w_I$ are larger, both $S_0$ and $1-S_1$ will be of order $1$. This is illustrated in the right  panel of Fig.~\ref{fig4}. While the region over which $\Lambda(S)$ is close to $1$ is smaller, resulting in a smaller entropy, it does not change significantly in the low-firing or high-firing cases, resulting in lower fragility.

\section*{Discussion}
Here we have shown that Shannon entropy of neural network dynamics is sensitive to the structure of excitatory and inhibitory interactions. Generally, high entropy is obtained by balancing E and I synaptic efficacy such that the system operates near the tipping point between two phases of network dynamics. Entropy is high all along this boundary, i.e., for a wide range of properly balanced E/I combinations. However, the regions within this boundary with the highest entropy are not robust; small variations in the synaptic strengths $W_E$, $W_I$, and in the fraction of inhibitory neurons $\alpha$ could cause entropy to plummet, drastically reducing the accessible states and disrupting the functioning of the network. We found that entropy is more robust when the effective synaptic strengths are larger. Given that $W_E$, $W_I$, and $\alpha$ are inevitably somewhat variable during development, across brain regions, and across individuals \cite{sahara, meinecke, hendry}, robustness to $W_E$, $W_I$, and  $\alpha$ variability may be important.
For networks constrained such that $W_E \sim W_I$  \cite{deneve, wehr, haider}, our findings imply that a small, nonzero fraction $\alpha >0$ of inhibitory neurons would result in a more robust network entropy.
Our results suggest that a population of organisms with reliable and high entropy brains requires that small, nonzero fraction of neurons be inhibitory, which is consistent with what exists in mammalian cortex \cite{sahara, meinecke, hendry}. 

Although high entropy is likely to be beneficial for certain functions of cerebral cortex, other functions might be better served by a low entropy condition.  For example, as discussed in the introduction, lower entropy might improve sensory signal processing by increasing the signal-to-noise ratio.  In this context, a small shift towards the lower firing side of the phase transition might be beneficial.  Such temporary shifts can occur due to neuromodulation; for example, attention is known to shift cortical dynamics towards a regime with smaller collective fluctuations \cite{harris2011}.  However, a shift towards the high firing regime or too large a shift towards the extremely inhibition-dominant regime would likely be bad for function.  Indeed, extreme deviation from well-balanced excitation and inhibition is implicated in a variety of brain disorders. For instance, when inhibition is sufficiently weak relative to excitation, seizures occur, as in epilepsy \cite{dichter}. Too much inhibition is associated with Down's syndrome \cite{fernandez}. Autism is also associated with imbalanced excitation and inhibition \cite{merzenich, nelson}, both in terms of abnormal numbers of inhibitory neurons and strengths of synapses \cite{gogolla}. Our work suggests that the dysfunction associated with these disorders may be, in part, due to abnormal entropy of cortical network dynamics.

If high entropy is a beneficial property for brain circuits, then the robust maximization of entropy could be a phenotypic target of evolution in the nervous system. Our results suggest that hitting this target requires neural circuits that include some inhibitory neurons and operate near the tipping point of a phase transition.

\section*{Acknowledgments}
Calculations were performed on Trestles at the Arkansas High Performance Computing Center, which is funded through multiple National Science Foundation grants and the Arkansas Economic Development Commission. 

%\nolinenumbers

\end{document}